\documentclass{article}[12pt,a4paper]

\usepackage[fontsize=11pt]{fontsize}
\usepackage[english]{babel}
\usepackage{indentfirst}
\usepackage[letterpaper,top=2cm,bottom=2cm,left=3cm,right=3cm,marginparwidth=1.75cm]{geometry}

\usepackage{amsmath}
\usepackage{bm}
\usepackage{graphicx}
\usepackage[colorlinks=true, allcolors=blue]{hyperref}
\usepackage[numbers,sort & compress]{natbib}
\usepackage{authblk}

\title{Linear analysis of flow mode transition triggered by finite-sized particles in Rayleigh-Bénard convection}
\author[1]{Dai Shi}
\affil[1]{Department of Mechanical Engineering, Osaka University, 2-1 Yamadaoka, Suita, Osaka 565-0871, Japan}

\begin{document}
\maketitle

\section{Introduction}
Particle-laden flow with thermally-driven convection involved is important to many fields, such as the application of nanofluids \cite{ref1, ref2, ref3, ref4} and particle-based solar collectors \cite{ref5, ref6, ref7}, where the particles are known to have a significant effect on the entire flow and heat transfer process. 
Rayleigh-Bénard (RB) convection, where the flow is heated from below, is a classical and typical system to study thermally-driven convection. 
Oresta et al. \cite{ref8, ref9} studied the RB systems suspended with particles and found that the heat source contributed by the particles can promote the total Nusselt number (Nu), especially when the particle diameter is small. As particle size increased, this effect caused by the coupled particle heat became less important.
Park et al. \cite{ref10} addressed the importance of the coupled particle heat brought by point particles. 
A significant enhancement of Nu was found to occur with the increase in the particle heat capacity. This enhancement was shown to be pronounced enough that even the attenuation effect caused by the coupled particle dynamics on Nu can be overwhelmed.
Prakhar and Prosperetti \cite{ref11} established a linear theory to study the influence of small particles on fluid instabilities and focused on the perturbation problem. 
However, the above studies all focused on the so-called point-particle model, in which particles are approximated as points and the temperature inside every particle is assumed uniform. In realistic applications like fluidized bed reaction, treating particles as points may cause overprediction of the real Nu or particle meltdown \cite{ref12,ref13}, which calls for studies with a focus on finite-sized particles. 

Ardekani et al. \cite{ref14} focused on the temperature gradient inside the particle, and studied the effect of particle size on laminar pipe flows. A considerable heat transfer enhancement by adding large particles was observed. A similar effect of large particles was also reported in \cite{ref15}, showing the significant effect of finite-sized particles on the heat transfer process in particle-laden flows.
Takeuchi et al. \cite{ref16} predicted a regular oscillation flow motion inside the laminar RB system laden with finite-sized particles, where the heat conductivity ratio of particle to fluid is shown to be critical in triggering this special phenomenon. 
Despite the above studies on finite-sized particles, theoretical models on the flow evolution under the thermal effect of finite-sized particles are rare.

In this work, a theoretical model combining the flow-scale momentum equation with the particle-scale
heat exchange equation is established to study the flow evolution of the RB system laden with finite-sized particles. The temperature change inside the particle is solved, and the relationship between the flow mode transition and the inter-phase heat exchange process is explained.

\section{Model Setup}
\subsection{Averaging process for flow evolution}
\begin{figure}[htbp]
\centering
\includegraphics[width=1\textwidth]{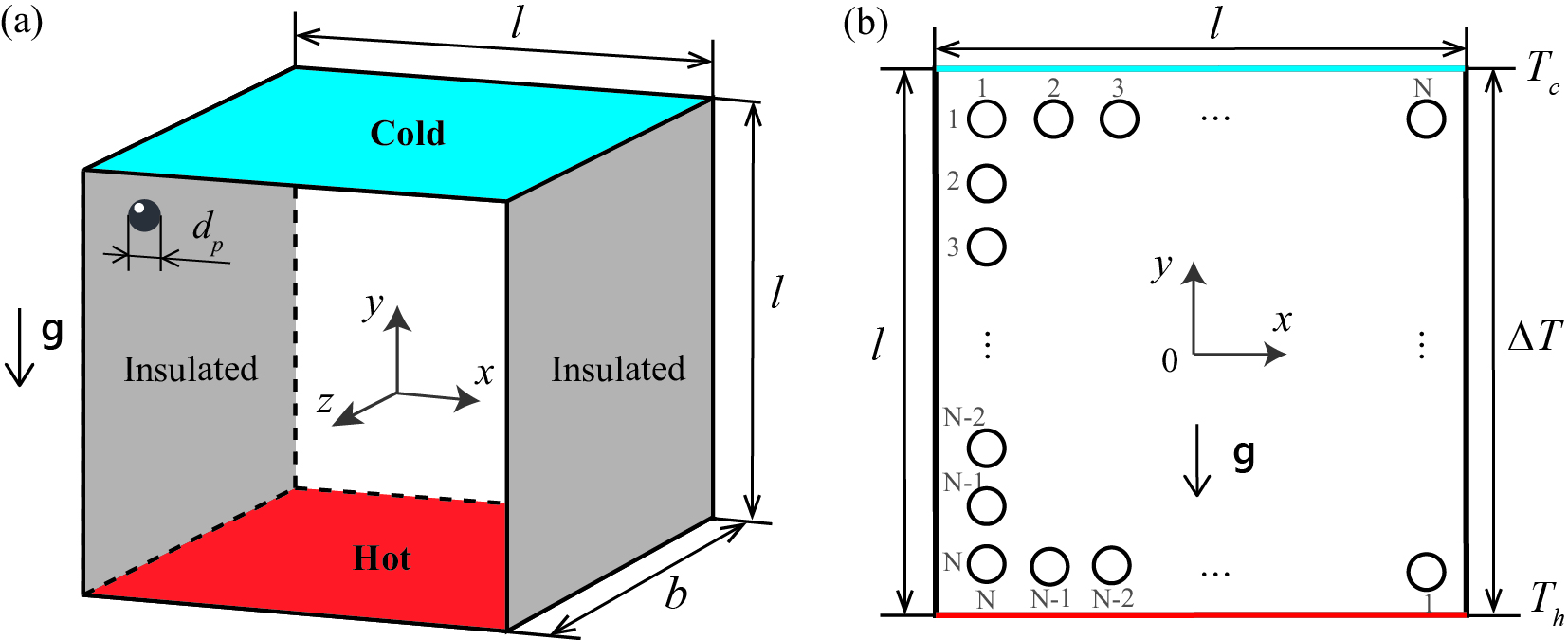}
\caption{\label{F1} The particle suspended RB convection (a) three-dimensional schematic (b) cross-section on the $x \mbox{-} y$ plane }
\end{figure}

An infinitely extended laminar system of RB convection along the $z$ direction, where the gravitational acceleration $\bm{g}$ acts in the $-\bm{e_y}$ direction, is considered in this study, as shown in Fig. \ref{F1}(a).
The cross-section of the system is a closed square cell of length $l$, with uniformly suspended spherical particles of diameter $d_p$ in both $x$ and $y$ direction, as shown in Fig. \ref{F1}(b).
The total particle number in the square cell is $N^2$.
The lateral walls of the system are thermally insulated,
and the temperature difference $\Delta T = T_h - T_c$ between the top and bottom plates is constant.
Throughout this study, the following properties of the fluid and particles are regarded as constant: density $\rho$, kinematic viscosity $\nu$, thermal conductivity $\lambda$, heat diffusivity $\kappa$, volumetric thermal expansion coefficient $\beta$, and specific heat $c$. 
The subscripts $f$ and $p$ indicate the fluid and particle phases, respectively. 

\begin{figure}[htbp]
\centering
\includegraphics[width=1\textwidth]{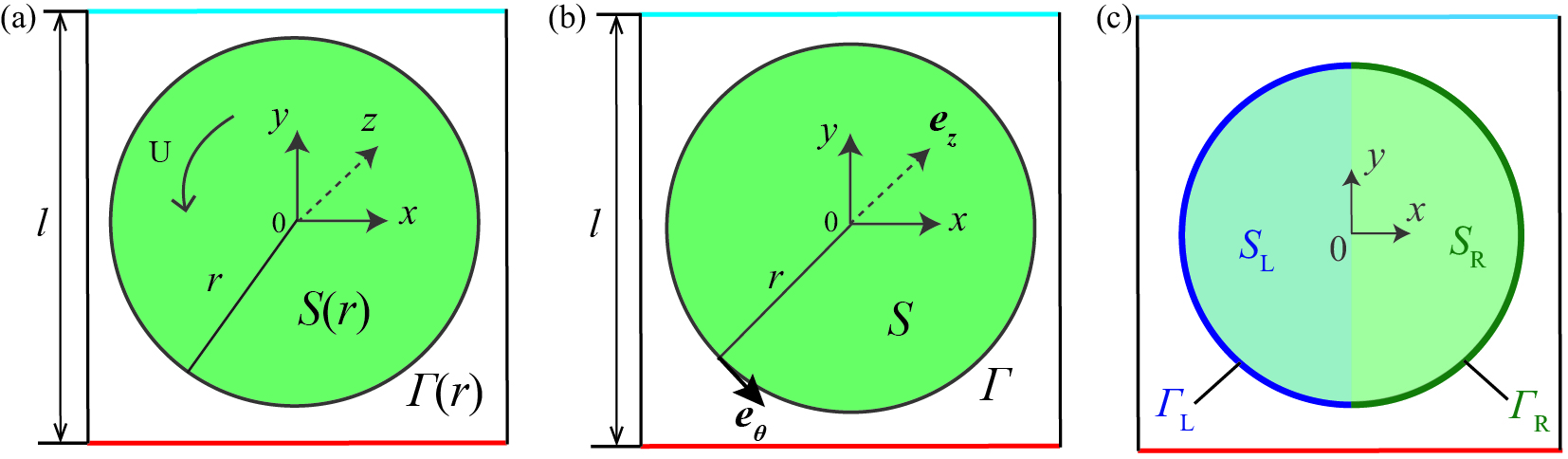}
\caption{\label{F2} Averaging process on the $x \mbox{-} y$ plane (a) the control volume of $S$ (b) the oriented surface (c) symmetrical parts of $S$ and $\Gamma$}
\end{figure}

Neglecting the impact of particle dynamics on the flow motion, the nondimensional equation for the time evolution of vorticity can be given as:
\begin{equation}
    \frac{\partial}{\partial t^*} \left ( \nabla^* \times \bm{u}^* \right ) + \nabla^* \times \left [ (\bm{u}^* \cdot \nabla^*) \bm{u}^* \right] = \sqrt{\frac{{\rm Pr}}{{\rm Ra}}} \nabla^* \times (\nabla^{*2} \bm{u}^*) + \nabla^* \times (T_f^* \bm{e_y}),
\end{equation}
where 
\begin{math}
  {\rm{Pr}} = \frac{\nu_f}{\kappa_f}
\end{math} is the Prandtl number,  
\begin{math}
  {\rm{Ra} }= \frac{\beta g \Delta T l^3}{\nu_f \kappa_f} 
\end{math} is the Rayleigh number,
$\bm{u}$ is the fluid velocity, and $T_f$ is the fluid temperature. 
In the above, $\Delta T$ is the characteristic temperature, and $t_{r1} = l/U $ is set as the reference time, where
\begin{math}
    U= \sqrt{g \beta \Delta T l}
\end{math}.
A coherent convective flow with negligible radial velocity is assumed to rotate around the domain center of the closed container in the cross-section and infinitely extends in the $z$ direction. 
Focusing on the $x \mbox{-} y$ plane, the continuity equation can be simplified to:
\begin{equation}
    \frac{\partial u_\theta^*}{\partial \theta}=0 ,
\end{equation}
where $\theta$ is the angular coordinate.
A circular control volume of $S(r)$ bounded by $\Gamma(r)$ is supposed to cover the convective flow, as shown in Fig. \ref{F1}(a), where $r$ is the radial coordinate.
Based on $S$, an oriented surface with the surface element vector of $\bm{e}_z d S$ can be defined. This oriented surface is enclosed by an oriented curve, whose unit vector is $\bm{e}_{\theta}$, as shown in Fig. \ref{F2}(b).
Integrating Eq. (1) over this oriented surface gives:
\begin{equation}
    \frac{\partial}{\partial t^*} \langle u_{\theta}^* \rangle_\Gamma = \sqrt{ \frac{{\rm{Pr}}} {{\rm{Ra}}} } \left ( \frac{\partial^2}{\partial r^{*2}} \langle u_{\theta}^* \rangle_\Gamma + \frac{1}{r^*}\frac{\partial}{\partial r^*} \langle u_{\theta}^* \rangle_\Gamma  \right ) +
    \frac{1}{2} \left ( 
    \langle  T_f^* \rangle_{\Gamma R} - \langle  T_f^* \rangle_{\Gamma L}  \right ),
\end{equation}
where $\langle \cdot \cdot \cdot \rangle _ {\Gamma}$ denotes the curve average.
In the above, $\langle u_\theta \rangle_\Gamma $ is the curve-averaged flow velocity, which comes from:
\begin{equation}
    \int_s (\nabla \times \bm{u}) \cdot \bm{e_z} dS = \int_\Gamma \bm{u} \cdot \bm{e_\theta} d \Gamma = \int_\Gamma u_\theta d \Gamma = \Gamma \langle u_\theta \rangle_\Gamma .
\end{equation}
The convective term is simplified to zero with the continuity equation:
\begin{equation}
    \int_s \nabla \times [ (\bm{u} \cdot \nabla) \bm{u} ] \cdot  \bm{e_z} dS = \int_0^{2 \pi} r \left ( 
    u_r \frac{\partial u_\theta}{\partial r} + \frac{u_\theta}{r} \frac{\partial u_\theta}{\partial \theta}
    \right) d \theta = 0 , 
\end{equation}
where $u_r$ is the radial velocity.
The first term on the right-hand side of Eq. (3) comes from the viscosity term:
\begin{equation}
    \int_s \nabla \times (\nabla^2 \bm{u}) \cdot \bm{e_z} dS
    = \int_0^{2 \pi} r \left(
    \frac{\partial^2 u_\theta}{\partial r^2} + \frac{1}{r} \frac{\partial u_\theta}{\partial r} + \frac{1}{r^2} \frac{\partial^2 u_\theta}{\partial \theta^2}\right ) d \theta = \Gamma \left ( \frac{\partial^2}{\partial r^2} \langle u_\theta \rangle_\Gamma + \frac{1}{r} \frac{\partial}{\partial r} \langle u_\theta \rangle_\Gamma    \right ),
\end{equation}
which describes the momentum diffusion along the radial direction.
The last term on the right-hand side of Eq. (3) comes from the buoyancy term:
\begin{equation}
    \int_s \nabla \times (T_f \bm{e_y}) \cdot \bm{e_z} dS = \int_s \frac{\partial T_f}{\partial x} \bm{e_z} \cdot \bm{e_z} dS = 
    \frac{\Gamma}{2} \left ( \langle T_f \rangle_{\Gamma R} - \langle T_f \rangle_{\Gamma L}     \right ) ,
\end{equation}
where $S = S_L + S_R$ and $\Gamma = \Gamma_L + \Gamma_R$, as shown in Fig. \ref{F2}(c). Eq. (3) shows that the imbalance of averaged fluid temperature between the left and right sides of the system is the triggering factor to invoke the bulk flow from rest to rotation.

\begin{figure}[htbp]
\centering
\includegraphics[width=1\textwidth]{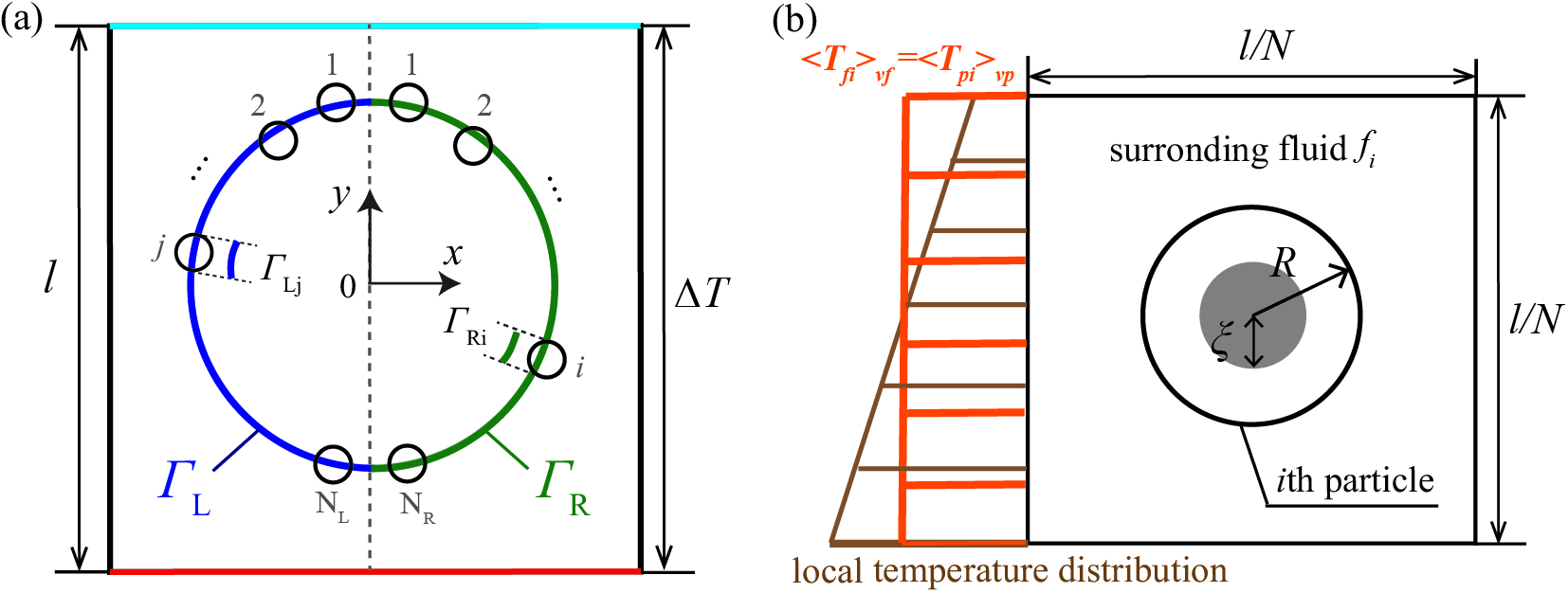}
\caption{\label{F3} Particle distribution on $\Gamma$ (a) symmetric particle distribution (b) the thermal equilibrium state of $\textit{i}$th particle and its surrounding fluid}
\end{figure}

The thermal impact of particles on flow evolution through the inter-phase heat exchange process is focused on in this study. Assuming a linear condition where the convection is suppressed and the symmetric particle array is kept without particle-particle contact during the flow motion, 
the evolution of fluid temperature located on the right side of the system depends mainly on the fluid-particle heat exchange:
\begin{equation}
    \left( \rho_f c_{pf} \frac{\Gamma}{2}  \right) \frac{d}{dt} \langle T_f \rangle_{\Gamma R} = - \sum_{i=1}^{N_R} \rho_p c_{pp} \frac{d}{dt} \langle T_{pi} \rangle_{vp} \Gamma _{Ri} ,
\end{equation}
where $N_{R}$ is the total particle number distributed on $\Gamma_R$, $v_p$ is the volume of every single particle, $\langle T_{pi} \rangle_{vp}$ is the volume-averaged temperature of the \textit{i}th particle on $\Gamma_R$, and $\Gamma_{R i}$ is the corresponding fraction of $\Gamma_R$ occupied by the \textit{i}th particle, as shown in Fig. \ref{F3}(a).
Considering the flow motion is tiny with suppressed convection, we assume the \textit{i}th particle is always surrounded by the same fluid element $f_i$ of volume $v_f$ and exchanges heat only with this $f_i$, as shown in Fig. \ref{F3}(b), where $v_f = \left ( \frac{l}{N}\right )^3 - v_p $. 
Therefore, the temperature evolution of the \textit{i}th particle can be written as:
\begin{equation}
     m_{p} c_{pp} \frac{d}{d t} \langle T_{pi} \rangle_{vp} = \pi d_p^2 h_p \left [ \langle T_{fi} \rangle_{vf} - \langle T_{pi} \rangle_{vp} \right ] ,
\end{equation}
where $m_{p}$ is the particle mass, $h_p$ is the particle heat transfer coefficient, and $\langle T_{fi} \rangle_{vf}$ is the volume-averaged temperature of the fluid around the \textit{i}th particle.
Combining with Eq. (8), we have:
\begin{equation}
    \frac{\Gamma}{2} \frac{d}{dt} \langle T_f \rangle_{\Gamma R} = \frac{1}{H \tau_{th}} \sum_{i=1}^{N_R} \left [ \langle T_{pi} \rangle_{vp} - \langle T_{fi} \rangle_{vf} \right ] \Gamma _{Ri},
\end{equation}

\begin{equation}
    H = \frac{\rho_f c_{pf}}{\rho_p c_{pp}} ,
\end{equation}

\begin{equation}
    \tau_{th} = \frac{m_p c_{pp}}{\pi d_p^2 h_p} = \frac{1}{6 {\rm{Nu_p}}} \frac{\lambda_p}{\lambda_f} \frac{d_p^2}{\kappa_p} ,
\end{equation}
where the particle Nusselt number ${\rm{Nu_p}}$ comes from the dimensionless process of $h_p$. The empirical correlation for ${\rm Nu_p}$ in natural convection systems with immersed spheres can be employed \cite{ref17, ref18}.
Similarly, the temperature evolution of $\langle T_f \rangle_{\Gamma L}$ can be given as:
\begin{equation}
     \frac{\Gamma}{2} \frac{d}{dt} \langle T_f \rangle_{\Gamma L} = \frac{1}{H \tau_{th}} \sum_{j=1}^{N_L} \left [ \langle T_{pj} \rangle_{vp} - \langle T_{fj} \rangle_{vf} \right ] \Gamma _{Lj},
\end{equation}
where $N_L$ is the total particle number distributed on $\Gamma_L$, $\langle T_{pj} \rangle_{vp}$ is the volume-averaged temperature of the \textit{j}th particle on $\Gamma_L$, and $\Gamma_{L j}$ is the corresponding fraction of $\Gamma_L$ occupied by the \textit{j}th particle.
Using Eq. (3), (10), and (13), we have:
\begin{equation}
\begin{aligned}
   \frac{\partial^2}{\partial t^{*2}} \langle u_{\theta}^* \rangle_\Gamma 
    & = \sqrt{ \frac{{\rm{Pr}}} {{\rm{Ra}}} } \frac{\partial}{\partial t^*} \left ( \frac{\partial^2}{\partial r^{*2}} \langle u_{\theta}^* \rangle_\Gamma + \frac{1}{r^*}\frac{\partial}{\partial r^*} \langle u_{\theta}^* \rangle_\Gamma  \right ) \\
    & + \frac{1}{\Gamma} \frac{1}{H St_{th}}
    \left [ 
    \sum_{i=1}^{N_R} \left ( \langle T_{pi}^* \rangle_{vp} - \langle T_{fi}^* \rangle_{vf} \right ) \Gamma _{Ri} - \sum_{j=1}^{N_L} \left ( \langle T_{pj}^* \rangle_{vp} - \langle T_{fj}^* \rangle_{vf} \right ) \Gamma _{Lj}
    \right ], 
\end{aligned}
\end{equation}
where $St_{th} = \tau_{th} / t_{r1} $ is the non-dimensional time of thermal relaxation. Therefore, the flow velocity can be regarded as being updated every time interval $\Delta t^* = H St_{th}$ by the inter-phase heat exchange process, which happens at the particle scale.

\subsection{Particle-scale heat exchange process}
\begin{figure}[htbp]
\centering
\includegraphics[width=0.5\textwidth]{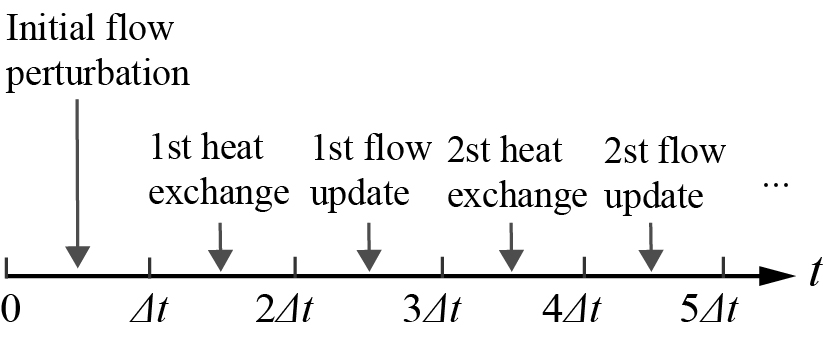}
\caption{\label{F4} The update process of heat exchange and flow evolution}
\end{figure}

In this study of linear analysis, the time scale of $t_{r2} = d_p / U$ is defined in the particle scale.
The mutually updated process of heat transfer and flow motion can be illustrated in Fig. \ref{F4}, where $\Delta t = t_{r2} \Delta t^* $.

\begin{itemize}
    \item 
$t=0$: Particles and the fluid rest in thermal equilibrium.
    \item 
$0<t< \Delta t$: Fluid moves with an initial perturbation velocity $\langle u_\theta^* \rangle_\Gamma$, resulting in a displacement of $l_A \equiv \langle u_\theta^* \rangle_\Gamma U \Delta t$ away from the thermal equilibrium position.
     \item 
$t= \Delta t$ : For every particle on the right system, $ \langle T_{pi} \rangle_{vp} - \langle T_{fi} \rangle_{vf} \equiv - \frac{\Delta T}{l} l_A = - T_A $. For every particle on the left system, $ \langle T_{pj} \rangle_{vp} - \langle T_{fj} \rangle_{vf} \equiv \frac{\Delta T}{l} l_A = T_A $. 
     \item 
$\Delta t <t< 2 \Delta t$: Heat exchange between every particle and its surrounding fluid begins.
     \item 
$t = 2 \Delta t$: $ \langle T_{pi} \rangle_{vp} - \langle T_{fi} \rangle_{vf}$ and $ \langle T_{pj} \rangle_{vp} - \langle T_{fj} \rangle_{vf}$ are updated as the value at $t = 2 \Delta t$.
     \item 
$ 2 \Delta t < t < 3 \Delta t$: Fluid moves with an updated velocity that satisfies Eq. (14).
\end{itemize}
Considering the symmetric particle arrangement on $\Gamma$ is kept, Eq. (14) can be rewritten as:
\begin{equation}
    \frac{\partial^2}{\partial t^{*2}} \langle u_\theta^* \rangle_\Gamma = \frac{\Gamma_p}{\Gamma} \frac{1}{H St_{th}} \left (  
    \langle T_{pi}^* \rangle_{vp} - \langle T_{fi}^* \rangle_{vf}
    \right ),
\end{equation}
where $\Gamma_p$ refers to the fraction of $\Gamma$ that occupied by the total particles distributed on $\Gamma$. 
Taking $t = \Delta t$ as the base state, it is suggested from the above process that the flow evolves in a way affected by the heat exchange process, where three time scales are involved:
\begin{equation}
    \Delta t = \frac{\rho_f c_{pf}}{\rho_p c_{pp}} \frac{1}{6 {\rm{Nu_p}}} \frac{\lambda_p}{\lambda_f} \frac{d_p^2}{\kappa_p} \frac{d_p}{l} ,
\end{equation}
\begin{equation}
    \tau_p = \frac{d_p^2}{4 \kappa_p} ,
\end{equation}
\begin{equation}
    \tau_f = \frac{d_f^2}{4 \kappa_f} , \ d_f = \left ( \frac{6}{\pi} v_f \right )^{1/3} .
\end{equation}
In this study, $ d_p / l \sim 10^{-2}$ is assumed and $\tau_f > \Delta t$ is indicated. Therefore, the $ \Delta t$ time interval is always insufficient for heat to update through the fluid region of $d_f$.
Considering our interest in high-thermal-conductivity particles, a precondition of $\tau_f \gg \tau_p$ is further assumed, and three regimes can be defined:
\begin{itemize}
    \item 
I. $\tau_p > \Delta t$ and $\tau_f > \Delta t$, which means $\frac{\rho_f c_{pf}}{\rho_p c_{pp}} \frac{\lambda_p}{\lambda_f} < P$;
    \item
II. $\tau_p < \Delta t < \tau_f $, which means $\frac{\rho_f c_{pf}}{\rho_p c_{pp}} \frac{\lambda_p}{\lambda_f} > P$;
    \item
III. $\tau_p \sim \Delta t$ and $\tau_f > \Delta t$, which means $\frac{\rho_f c_{pf}}{\rho_p c_{pp}} \frac{\lambda_p}{\lambda_f} \sim P$.  
\end{itemize}
In the above,
\begin{equation*}
    \frac{\tau_p}{\Delta t} = \frac{\rho_p c_{pp}}{\rho_f c_{pf}} \frac{\lambda_f}{\lambda_p} \frac{l}{d_p} \frac{3 {\rm{Nu_p}}}{2} = \frac{\rho_p c_{pp}}{\rho_f c_{pf}} \frac{\lambda_f}{\lambda_p} P .
\end{equation*}
Heat exchange methods vary in different regimes, resulting in different flow modes.

\section{Results and Discussion}
The former analysis has given significance to the time scales involved in the heat exchange process. In this section, detailed temperature evolution in both particle and fluid regions, which varies with the time scales, is discussed. Considering the precondition of $\tau_f \gg \tau_p$, the particle temperature change can be treated as a Cauchy problem.

\subsection{Flow mode I}
\begin{figure}[htbp]
\centering
\includegraphics[width=1\textwidth]{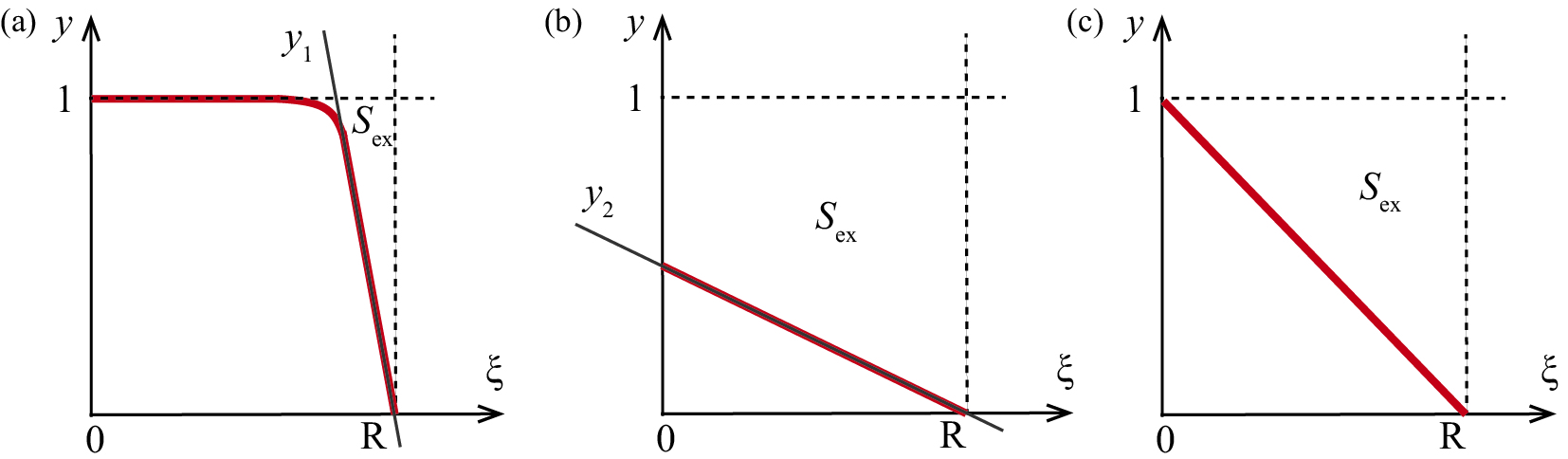}
\caption{\label{F5} Temperature change inside the particle (a) Regime I (b) Regime II (c) Regime III}
\end{figure}

In Regime I, the thermal distance traversed within the particle region during the $\Delta t$ time interval is:
\begin{equation*}
    l_p = \sqrt{\kappa_p \Delta t} < R ,
\end{equation*}
where $R=0.5 d_p$. Describing the thermal base state  as:
\begin{equation}
     \langle T_{pi} \rangle_{vp, t=\Delta t} - \langle T_{fi} \rangle_{vf, t=\Delta t} = T_{p0} - T_{f0} = - T_A ,
\end{equation}
thermal state at the end of the first heat exchange process can be written as:
\begin{equation}
     \langle T_{pi} \rangle_{vp, t=2\Delta t} - \langle T_{fi} \rangle_{vf, t=2\Delta t} = T_{p0} + \Delta T_p - T_{f0} - \Delta T_f ,
\end{equation}
where $\Delta T_p$ refers to the volume-averaged temperature change of the single particle located on $\Gamma_R$ during $\Delta t < t < 2\Delta t$, and $\Delta T_f$ refers to the volume-averaged temperature change of the corresponding fluid element.
Under the initial inter-phase temperature difference, heat is transferred inside the particle region along the radial direction, which can be simplified as a one-dimensional heat conduction problem.
As a result, $\Delta T_p$ can be written as:
\begin{equation}
    \begin{split}
        \Delta T_p 
           &= \frac{1}{2R} \int_{-R}^R d \xi \int_{- l_p}^{l_p} T_A  \frac{{\rm{exp}} \left [ - \frac{(\xi - \xi_0)^2}{4 \kappa_p \Delta t}\right ] } {\sqrt{4 \pi \kappa_p \Delta t}} d \xi_0 
        \\ &= \frac{T_A}{2R \sqrt{4 \pi \kappa_p \Delta t}} 
            \frac{l_p}{R} \int_{-R}^R d \xi \int_{-R}^R {\rm{exp}}\left [- \frac{(\xi - \xi_0)^2}{4 \kappa_p \Delta t} \right ] d \xi_0 ,
    \end{split}
\end{equation}
where $\xi$ is the integration variable referred to the existing region of every single particle on the $x \mbox{-} y$ plane, as shown in  Fig. \ref{F3}(b), and $\xi_0$ is the integration variable referred to $l_p$.
The dual integral part in Eq. (21) can be rewritten with a transform of the integration variable:
\begin{equation}
\begin{split}
        \int_{-R}^R d \xi \int_{-R}^R {\rm{exp}}\left [- \frac{(\xi - \xi_0)^2}{4 \kappa_p \Delta t} \right ] d \xi_0
        &= \int_{-R}^R d \xi \int_{\xi + R}^{\xi -R }  {\rm{exp}} \left [  - \frac{\eta ^2}{4 \kappa_p \Delta t}                 \right ] d \eta
        \\
        &= \int_{0}^R d \xi    \left (  \int_{\xi+R}^{-(\xi+R)}  d \eta - \int_{-(\xi-R)}^{\xi-R}  d \eta \right )  {\rm{exp}} \left [- \frac{\eta ^2}{4 \kappa_p \Delta t} \right ]
        \\ 
        &\approx \sqrt{4 \pi \kappa_p \Delta t} \left ( R - \int_0^R \sqrt{ 1- {\rm{exp}} \left [- \frac{(R- \xi) ^2}{4 \kappa_p \Delta t} \right ]} d \xi \right ) ,
\end{split}
\end{equation}
where $\eta = \xi - \xi_0$ . Combining with Eq. (21), we have:
\begin{equation}
    \Delta T_p = \frac{T_A}{2R} \frac{l_p}{R} \left ( R - \int_0^R \sqrt{ 1- {\rm{exp}} \left [- \frac{(R-\xi) ^2}{4 \kappa_p \Delta t} \right ]} d \xi \right ) = \frac{T_A}{2R} \frac{l_p}{R} S_{ex} ,
\end{equation}
where $S_{ex}$ represents the integral area enclosed by $\xi = R$, $y =1$, and $ y= \sqrt{ 1- {\rm{exp}} \left [- \frac{(R - \xi) ^2}{4 \kappa_p \Delta t} \right ]}$, the general plot of which is shown as the red curve in Fig. \ref{F5}(a).
The linear part of the red curve can be similarized to fit $y_1 = C_1 \frac{R - \xi}{\sqrt{4 \kappa_p \Delta t}}$, leading to an approximation of $S_{ex}$:
\begin{equation}
    S_{ex} = \frac{1}{2} \times \frac{\sqrt{4 \kappa_p \Delta t}}{C_1} \times 1 = \frac{\sqrt{\kappa_p \Delta t}}{C_1} , 
\end{equation}
where $C_1$ is a constant coefficient. Combining with Eq. (23), we have:
\begin{equation}
    \Delta T_p = T_A \frac{l_p \sqrt{\kappa_p \Delta t} }{2 C_1 R^2} \sim T_A \frac{\kappa_p}{\kappa_f} \frac{d_p}{l} ,
\end{equation}

\begin{equation}
    \Delta T_f = - \frac{1}{HE} \Delta T_p \sim   - T_A \frac{\lambda_p}{\lambda_f} \frac{d_p}{l} ,
\end{equation}
where $E= v_f / v_p$. Combining with Eq. (20), we have:
\begin{equation}
    \langle T_{pi} \rangle_{vp, t=2\Delta t} - \langle T_{fi} \rangle_{vf, t=2\Delta t} \sim
    - T_A \left (1- \frac{\kappa_p}{\kappa_f} \frac{d_p}{l} -  \frac{\lambda_p}{\lambda_f} \frac{d_p}{l} \right ) .
\end{equation}
Using Eq. (15), the flow motion during $ 2 \Delta t < t < 3 \Delta t$ obeys the following relationship:
\begin{equation}
     \frac{\partial^2}{\partial t^{*2}} \langle u_\theta^* \rangle_\Gamma = - \frac{\Gamma_p}{\Gamma} \frac{d_p}{l} 
     \left (  
     1- \frac{\kappa_p}{\kappa_f} \frac{d_p}{l} -  \frac{\lambda_p}{\lambda_f} \frac{d_p}{l}
    \right )  \langle u_\theta^* \rangle_\Gamma  ,
\end{equation}
which indicates a flow mode of simple harmonic (SH) motion, oscillating around the thermal equilibrium position.

\subsection{Flow mode II}
In Regime II, the heat transfer process through the particle region should be finished for at least one round during the time interval of $\Delta t$, which means:
\begin{equation}
        \Delta T_p 
           = \frac{1}{2R} \int_{-R}^R d \xi \int_{- R}^{R} T_A  \frac{{\rm{exp}} \left [ - \frac{(\xi - \xi_0)^2}{4 \kappa_p \Delta t}\right ] } {\sqrt{4 \pi \kappa_p \Delta t}} d \xi_0 
          = \frac{T_A}{2R} S_{ex} ,
\end{equation}

\begin{equation}
    S_{ex} = R \times 1 - \frac{1}{2} \times R \times \frac{R C_2}{\sqrt{4 \kappa_p \Delta t}} 
    = R - \frac{R^2 C_2}{4 \sqrt{\kappa_p \Delta t}} . 
\end{equation}
In the above, $S_{ex}$ is the trapezoidal area enclosed by $\xi = R$, $\xi = 0$,  $y =1$, and $ y= \sqrt{ 1- {\rm{exp}} \left [- \frac{(R - \xi) ^2}{4 \kappa_p \Delta t} \right ]}$, the linear part of which is fitted to $y_2 = C_2 \frac{R - \xi}{\sqrt{4 \kappa_p \Delta t}}$, as shown in Fig. \ref{F5}(b). 
Therefore, the temperature change during the time interval can be written as:
\begin{equation}
    \Delta T_p = T_A \left ( \frac{1}{2} - \frac{R C_2}{8 \sqrt{\kappa_p \Delta t}}  \right ) 
    \sim   T_A \left ( \frac{1}{2} - \sqrt{\frac{\kappa_f}{\kappa_p} \frac{l}{d_p} } \right ) ,
\end{equation}
\begin{equation}
    \Delta T_f = - \frac{1}{HE} \Delta T_p 
    \sim   - T_A \left ( \frac{1}{2HE} - \frac{\lambda_p}{\lambda_f} \frac{\kappa_f}{\kappa_p} \sqrt{ \frac{\kappa_f}{\kappa_p} \frac{l}{d_p}  } \right ) .
\end{equation}
Combining with Eq. (15) and Eq. (20), the flow motion during $ 2 \Delta t < t < 3 \Delta t$ obeys the following relationship:
\begin{equation}
     \frac{\partial^2}{\partial t^{*2}} \langle u_\theta^* \rangle_\Gamma = 
     - \frac{\Gamma_p}{\Gamma} \frac{d_p}{l} 
     \left (  
     \frac{1}{2}- \frac{1}{2HE}  + \sqrt{ \frac{\kappa_f}{\kappa_p} \frac{l}{d_p} } +  \frac{\lambda_p}{\lambda_f} \frac{\kappa_f}{\kappa_p} \sqrt{\frac{\kappa_f}{\kappa_p} \frac{l}{d_p} }  \right )  
     \langle u_\theta^* \rangle_\Gamma ,
\end{equation}
which also indicates a flow mode of SH motion around the thermal equilibrium position.

\subsection{Flow mode III}
After the above discussion, the flow pattern in Regime III, where $\Delta t$ is about the time to complete one round of heat transfer through the particle region, becomes easier to understand.
As shown in Fig. \ref{F5}(c), the averaged temperature change can be given as:
\begin{equation}
    \Delta T_p 
           = \frac{T_A}{2R} S_{ex} = \frac{T_A}{4},
\end{equation}

\begin{equation}
     \Delta T_f = - \frac{1}{HE} \Delta T_p = - \frac{T_A}{4HE} .
\end{equation}
As a result, flow motion during $ 2 \Delta t < t < 3 \Delta t$ obeys the following relationship:
\begin{equation}
    \frac{\partial^2}{\partial t^{*2}} \langle u_\theta^* \rangle_\Gamma = 
     - \frac{\Gamma_p}{\Gamma} \frac{d_p}{l} 
     \left ( \frac{3}{4} - \frac{1}{4HE}  \right ) \langle u_\theta^* \rangle_\Gamma ,
\end{equation}
which indicates the flow period has no obvious connection with the thermal conductivity ratio.

Regime III occurs only when $\sqrt{\kappa_p \Delta t} \approx R$, which means it is easy to shift to Regime I or Regime II.
The previous discussion shows that with the increase of $\lambda_p / \lambda_f$, the oscillation period in Regime I becomes shorter while the oscillation period in Regime II becomes longer.
Therefore, there should exist a SH flow mode in Regime III, which yields the minimum oscillation period.

\subsection{Particle motion}
Considering the symmetric particle array is always kept in this study, it is reasonable to assume that particles distributed on $\Gamma$ have the same velocity:
\begin{equation*}
    u_{pi, \Gamma} = u_{p, \Gamma} = u_{p \theta, \Gamma} ,
\end{equation*}
where $u_{p \theta, \Gamma}$ is the particle velocity along the angular direction.
Therefore, the drag force experienced by every single particle on $\Gamma$ can be written as:
\begin{equation}
    f_d = m_p \frac{\langle u_\theta \rangle_\Gamma - u_{p \theta, \Gamma}}{\tau_m} ,
\end{equation}
where a positive $f_d$ indicates the drag force acts along the counterclockwise direction, and a negative $f_d$ indicates the drag force acts along the clockwise direction.
In the above, $\tau_m$ is the particle relaxation time, which can be written in the general form\cite{ref19}:
\begin{equation}
    \frac{1}{\tau_m} = \frac{3}{4} C_{D} f(\epsilon)^m \frac{\rho_f}{\rho_p} \epsilon^2 \frac{\left |\langle u_\theta \rangle_\Gamma - u_{p \theta, \Gamma} \right |}{d_p},
\end{equation}
where $C_{D}$ is the drag coefficient for single sphere suspension proposed by Schiller and Naumann \cite{ref20}, $\epsilon$ is the local void fraction or porosity, and $f(\epsilon)^m$ is the porosity function to correct the relaxation time when the particle suspension is not dilute enough. 

\begin{figure}[htbp]
\centering
\includegraphics[width=0.5\textwidth]{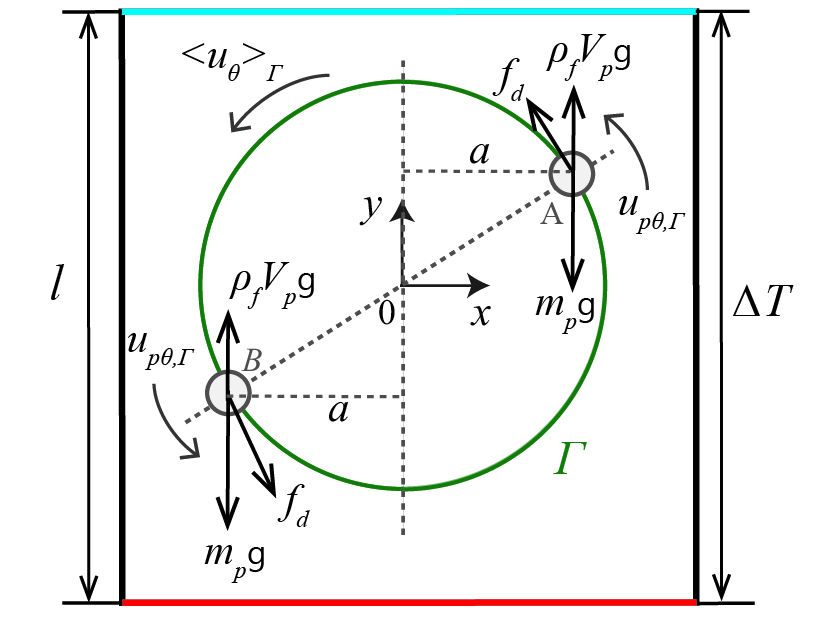}
\caption{\label{F6} The symmetric particle couple}
\end{figure}

Treating particles distributed on $\Gamma$ as symmetric particle couples, for example, particle $A$ and $B$ in Fig. \ref{F6}, the moment of force experienced by every particle couple can be written as:
\begin{equation}
    \sum M_{AB} = (\rho_{f} v_{p} g a - \rho_{f} v_{p} g a)+ (-m_{p} g a + m_{p} g a)+ 2  f_d r = 2 f_d r,
\end{equation}
suggesting the particle motion is mainly driven by the drag force. 
According to the conservation of angular momentum, we have:
\begin{equation}
    \frac{d}{dt} L_{p, \Gamma} = \frac{N_\Gamma}{2} I_p \frac{d w_{\Gamma}}{dt} = \frac{N_\Gamma}{2} \sum M_{AB} ,
\end{equation}
where $L_{p, \Gamma}$ represents the angular momentum of particles located on $\Gamma$, $N_\Gamma$ is the total particle number on $\Gamma$, $I_2$ is the moment of inertia of every particle couple, and $w_{\Gamma}$ is the angular velocity of every particle.
Assuming the particle movement is tiny enough that $
    w_{\Gamma} \approx u_{p \theta, \Gamma} $, Eq. (43) can be rewritten as:
\begin{equation}
    \frac{d}{dt} u_{p \theta, \Gamma} = \frac{2 r m_p}{I_p \tau_m} \left (  \langle u_\theta \rangle_\Gamma -  u_{p \theta, \Gamma}    \right ) ,
\end{equation}
which suggests that particles also move with a kind of regular oscillation mode driven by the flow, while a phase difference exists between the particle SH  and fluid SH motions.

\subsection{Validation}
An oscillatory granular flow, which is driven by the simple harmonic fluid motion, is predicted from the above analysis. In this section, validation is carried out by comparing the oscillation period suggested by the current study with that from related simulation research.

\begin{figure}[htbp]
\centering
\includegraphics[width=0.65 \textwidth]{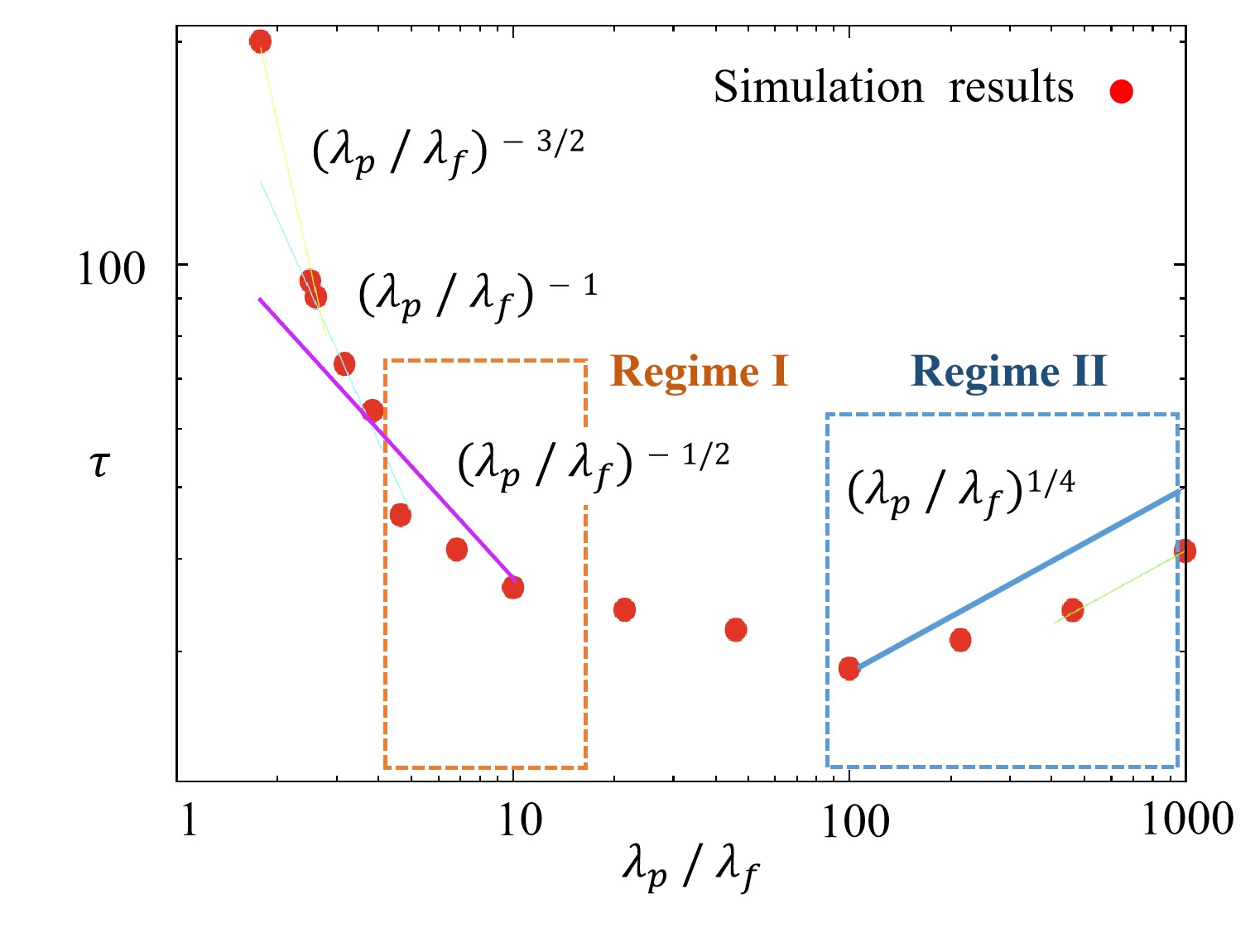}
\caption{\label{F7} Validation with simulation results}
\end{figure}

As shown in Fig. \ref{F7}, red dots refer to the simulation results of particle motion in the solid-dispersed Rayleigh-Bénard convection\cite{ref16}, where $d_p / l$ is set as 0.05 and $H$ is set as 1. Fitting these conditions, Regime I of this study falls into the range of 
$ \frac{\lambda_p}{\lambda_f} < 60 $ ,
and the granular flow yields a SH motion, whose period is:
\begin{equation}
    \tau_I = \left [ \frac{\Gamma_p}{\Gamma} \frac{d_p}{l} 
     \left (  
     1- H \frac{\lambda_p}{\lambda_f} \frac{d_p}{l} -  \frac{\lambda_p}{\lambda_f} \frac{d_p}{l}
    \right ) \right ] ^ {- \frac{1}{2}}
    \sim 
     (  \lambda_p / \lambda_f )^ {- \frac{1}{2}},
\end{equation}
showing agreement with the simulation study.
Regime II falls in the range of $ \frac{\lambda_p}{\lambda_f} > 60$ ,
where the granular flow yields a regular oscillation period:
\begin{equation}
    \tau_{II} = \left [ \frac{\Gamma_p}{\Gamma} \frac{d_p}{l} 
     \left (  \frac{1}{2}- \frac{1}{2HE}  + \sqrt{ \frac{\kappa_f}{\kappa_p} \frac{l}{d_p} } +  \frac{\lambda_p}{\lambda_f} \frac{\kappa_f}{\kappa_p} \sqrt{\frac{\kappa_f}{\kappa_p} \frac{l}{d_p} }  \right ) \right ] ^ {- \frac{1}{2}} 
     \sim \left ( \lambda_p / \lambda_f  \right )^{\frac{1}{4}} .
\end{equation}
This correlation between the oscillation period and thermal conductivity ratio shows especially good agreement with the simulation study.
Regime III, which is between Regime I and II, falls in the range around $ \frac{\lambda_p}{\lambda_f} = 60$, where a minimum oscillation period can be observed.
In the simulation results, the shortest oscillation period of all does occur around $\lambda_p / \lambda_f = 60 $ while the oscillation period still shows a relationship with the thermal conductivity ratio. 

It is also noted that the correlation in Regime I fails to cover the flow motion around the range of $ 1 < \frac{\lambda_p}{\lambda_f} <10 $, as indicated from the simulation results. The reason is supposed to be that the assumption of $\tau_f \gg \tau_p$ becomes invalid under the problem setting of $ \frac {\lambda_p}{\lambda_f} <10 $, and it is not scientific to simplify the particle temperature change into a Cauchy problem anymore. More heat transferred through the particle surface to the fluid element should be modeled.

\section{Conclusion}
This work starts with a mathematical model to study RB convection laden with finite-sized particles.
By employing an averaging method, the flow evolution after perturbation is focused on. A linear analysis is conducted to investigate the effect of the inter-phase heat exchange process on the flow mode transition.
It is shown that when a flow perturbation occurs, leading to an inter-phase temperature difference $T_A$, the flow velocity is updated by the heat exchange process every time interval of $\Delta t$, resulting in a regular oscillation flow motion.
The oscillation period is shown to be related to a group of physical quantities, which are determined by three time scales involved in the heat exchange process between every single particle and its surrounding fluid element. 
By assuming $\tau_f / \tau_p \gg 1$, when the thermal relaxation time $\tau_p$ for heat to reach one balance inside every single particle is longer than $\Delta t$, the flow oscillation period is shown to present a decreasing trend with the thermal diffusivity ratio $\kappa_p / \kappa_f $ of particle to fluid.
When the thermal relaxation time $\tau_p$ for heat to reach one balance inside every single particle is shorter than $\Delta t$, the flow oscillation period is shown to present an increasing trend with the thermal diffusivity ratio $\kappa_p / \kappa_f $.
Under the case of $\tau_p \sim \Delta t$, the shortest oscillation period is predicted to occur, the value of which does not vary significantly with the change of $\kappa_p / \kappa_f $.
Particle motion is also shown to be a regular oscillation, which is driven by the drag force to follow that of the flow motion, sharing a similar oscillation period with that of the flow. 

In this study, flow evolution is analyzed in detail for $ 2 \Delta t < t < 3 \Delta t$ after the heat exchange process during $ \Delta t < t < 2 \Delta t$, which starts after an initial inter-phase temperature difference at $ t = \Delta t$.
In fact, flow evolution during other time intervals is similar to that during $ 2 \Delta t < t < 3 \Delta t$, since the heat exchange process always starts after an initial inter-phase temperature difference of $T_A^{'}$:
\begin{equation}
    T_A^{'} = \frac{\Delta T}{l} l_A = \frac{\Delta T}{l} \langle u_\theta \rangle_\Gamma t_c \sim \frac{\Delta T}{l} \langle u_\theta \rangle_\Gamma \Delta t = T_A .
\end{equation}
Therefore, the flow mode (i.e. after $t = 3 \Delta t$ in this study) in the RB system laden with finite-sized high-thermal-conductivity particles is always a SH motion with time development, with changes only in the oscillation amplitude, affected by the varying $l_A$ at different time points.

However, the current study is based on the assumption of $\tau_f / \tau_p \gg 1$, which means the particle thermal diffusivity is much higher than the fluid thermal diffusivity. It becomes invalid to describe the flow evolution when the particle thermal diffusivity gets close to the fluid thermal diffusivity. 
Theoretical analysis under $\tau_f / \tau_p > 1$ is also a major ongoing work.

\section*{Acknowledgments}
This work was supported by JST SPRING, Grant Number JPMJSP2138.

\end{document}